# A new cut-based genetic algorithm for graph partitioning applied to cell formation

**Menouar Boulif**

*Department* of computer science. Faculty of sciences.
University M'hamed Bougara of Boumerdès.
Avenue de l'indépendance. 35000. Boumerdès, ALGERIA. {m.boulif@univ-boumerdes.dz}

**Abstract**: Cell formation is a critical step in the design of cellular manufacturing systems. Recently, it was tackled using a cut-based-graph-partitioning model. This model meets real-life production systems requirements as it uses the actual amount of product flows, it looks for the suitable number of cells, and it takes into account the natural constraints such as operation sequences, maximum cell size, cohabitation and non-cohabitation constraints. Based on this model, we propose an original encoding representation to solve the problem by using a genetic algorithm. We discuss the performance of this new GA in comparison to some approaches taken from the literature on a set of medium sized instances. Given the results we obtained, it is reasonable to assume that the new GA will provide similar results for large real-life problems.

## 1. Introduction

Cellular Manufacturing Systems (CMS) are an industrial implementation of the Group Technology (GT) philosophy. CMS consist of dividing the manufacturing system into cells so that similar parts are processed in the same cell. Such systems are specifically designed for job shops whose production volume is average (Boulif 2006). CMS have proven ability to reduce set-up times, in-process inventories, lot sizes and production equipment while improving productivity and production system mastery (Souilah 1994). There are four important steps in CMS design: (1) process planning, (2) cell formation (CF), (3) machine layout and (4) cell layout. Our paper deals with CF which is a key step in CMS design.

In the last decades, the interest of researchers on CF triggered a big amount of research that can be broadly divided into the following three non-exclusive categories (Boulif 2006):

**1. Methods based on the part-machine incidence matrix**: The part-machine incidence matrix (PMIM) is a binary matrix that indicates the set of machines used to process each part. A large number of studies concentrate on the use of this matrix, by considering that it is the most important, if not the sole, input of the problem (e.g. Nunkaew 2013). Such matrix-based methods generally proceed by swapping rows and/or columns of the PMIM to yield a diagonal block structure from which part families and machine cells are obtained. This category has several limitations as it takes neither the operation sequences nor the production volumes into account.

**2. Methods based on similarity coefficients**: McAuley (McAuley 1972) was the first to use the measure of similarity between machines to identify cells. He developed a mathematical coefficient that uses only PMIM information. Since his article was published, numerous papers have tried to enhance this measure by adding further inputs, including production volumes (Seifoddini 1989, Mahapatra 2008), part operational time and operation sequences (Gupta 1990). The efforts in this category tend to combine data inputs from several criteria, defining the similarity coefficient as a weighted combination of the overall criteria (see (Yin 2006) for a comprehensive study). However, weak justifications are given for the weighting procedure, which is an influential parameter in the derived solutions.

**3. Methods based on Meta-Heuristics**: CF problem's NP-completeness prompted research to focus on heuristic methods.

Meta-heuristics, have attracted the most attention, leading to Tabu search approach (Logendran 1995), Simulated Annealing (Venugopal 1992), Neural Network approaches (Kaparthi 1992) and Genetic algorithms (GA) (Venugopal 1992; Gupta 1995; Boulif 2006; Darla 2014). Literature findings proved that GA based methods are very interesting research paths in comparison to other heuristics (Mahapatra 2008). In GA based approaches, the encoding representation is the sole means that prospects the search space. We believe therefore that it must be lent more attention in research efforts. In fact, most of the published works (e.g. (Venugopal 1992; Gupta 1995; Darla 2014)) that use an evolutionary approach adopt the machine-to-cell integer encoding that has proven its limitation (Boulif 2006).

To contribute to these efforts, this paper proposes a *new* cut-based GA *encoding representation* derived from the cut-based-graph-partitioning model (Merchichi 2015). The proposed cut-based solving approach supposes that the number of cells is not known *a priori* and hence, it looks for the appropriate number of cells. Furthermore, the proposed approach is more suitable to meeting the real-life production systems requirements as it uses the actual amount of product flow that is falsely estimated by binary-PMIM-based methods, and as it considers the natural constraints such as operation sequences, maximum cell size, cohabitation and non-cohabitation constraints.

The remainder of this paper is organized as follows. In section 2, a graph partitioning formulation to the MCF problem is presented. Section 3 describes the proposed genetic algorithm. The next section discusses the results obtained by applying the proposed methods on some chosen data sets, and section 5 presents our conclusions as well as our recommendations for further research.

## 2. Formulation

In what follows we describe the MCF problem and present a graph theoretic model.

### 2.1 *Problem description*

Manufacturing cell formation consists of clustering machines into cells and part types into families. Each part family is dedicated to a machine cell, such that parts of the same family are essentially processed in their associated cell.

This clustering must meet the following practical constraints (Boulif 2006):

(1) an upper bound for cell size;
(2) cohabitation constraints requiring some machines to be placed in the same cell; and
(3) non-cohabitation constraints requiring the separation of some machine couples in different cells.

To have a good clustering, several criteria are considered in the literature. One of the most used is the minimization of intercellular part traffic.

When the two clustering tasks (part and machine clustering) are not done simultaneously, it is possible to deduce the part clustering from the machine clustering, and vice versa. Our approach focuses only on machine clustering into cells. The clustering of part types into families can be deduced in a second step, for instance, by assigning each part to the cell in which it spends the most of its production time.

### 2.2 *Mathematical model*

The following formulation models Cell Formation as a Graph Partitioning problem (Merchichi 2015).

*2.1.1 Input data*

(1) Let us consider $M = \{M_1, M_2, ..., M_m\}$ a set of $m$ machines and $P = \{P_1, P_2, ..., P_p\}$ a set of $p$ part types.
(2) For each part type $P_k$ ($k = 1, 2, ...., p$), we suppose given:
    (i) A single sequence (or routing) of machines to be visited by the part: $R_k = (\langle M_{k,1}\rangle, \langle M_{k,2}\rangle, ..., \langle M_{k,s_k}\rangle)$, where: $\langle M_{k,t}\rangle \in M$ $(t = 1,2, ..., s_k)$ and $s_k$ is the number of machines in the routing $R_k$.
    (ii) $r_k$: the mean production volume of part type $P_k$ per time unit.
(3) Constraint input data :
    (i) A set of machine couples $SC$.

(ii) Another set of machine couples $SN$. $SN \cap SC = \emptyset$ .
(iii) An integer number $N$.

*2.2.2 Flow graph construction*

(4) For each $(P_k, M_i, M_q) \in P \times M \times M$, we denote as $v_{k,i,q}$ the number of times $M_i$ follows $M_q$ or inversely $(i, q = 1, ..., m)$ in the routing of $P_k$ .

(5) For each couple $(M_i, M_q) \in M \times M$, we denote the $M_i$, $M_q$ inter-machine traffic $( i,q = 1, ..., m)$ by:

$$t_{iq} = \sum_{k=1}^{p} r_k \cdot v_{kiq}$$

We then define:

The non-oriented flow graph $G = (M, E)$, where the set of vertices $M$ is the set of machines and the set of edges $E$ is the set of non-ordered machine couples that are connected by a positive traffic or that are in $SC$ or $SN$:

$$E = \{e_{iq} \,/\, (M_i , M_q) \in M \times M,\ i,\ q = 1, ..., m;\ i \neq q \ \text{ and } \ t_{iq} \neq 0\} \cup SN \cup SC$$

(6) Edge weight function $W$:

$$W(e_{iq}) = t_{iq} \quad \text{where} \quad i , q \in \{1, ..., m \}.$$

*Remark* If the flow graph is not connected, it must be connected by adding fictive edges with null weights. This procedure permits the assumption that the flow graph is connected from here on.

*2.2.3 Decision variables*

(7) Let $CS=\{w_1, w_2, ..., w_{|CS|}\}$ be a subset of cuts of $G$ such that $\left[\frac{m}{N}\right] \leq |CS| < m$

where $[x]$ is the integer part of the number $x$ and |X| is the cardinal of the set X.

*2.2.4 Intermediate processing*

(8) Let $C = \{C_1, C_2, ..., C_J\}$ be the set of connected components of the graph

$$\tilde{G} = (M, E - \cup_{i=1}^{|CS|} w_i) .$$

$C$ is a partition of $M$ in $J$ cells. That is,

$$Cj \neq \emptyset,\ \forall\ j \in \{1,2, ...,J\} ;\ \bigcup_{j=1}^{J} C_j = M \ \text{ and } C_j \cap Cg = \emptyset,\ \forall\ j,g \in \{1,2, ...,J\},\ j \neq g.$$

(9) Subset of intercellular edges:

$$E\,(CS) = \cup_{i=1}^{|CS|} w_i$$

(10) Total intercellular traffic:

$$T(CS) = \sum_{e_{iq} \in E(CS)} W(e_{iq})$$

*2.2.5 Constraints*

To be feasible CS must respect the following constraints:

(1) Maximum number of machines allowed in a cell $N$: We only consider cut subsets $CS$ whose $C$ partition respects:

$$\forall\ C_j \in C\ (j = 1, 2, ...., J)\ \ |C_j| \leq N.$$

(2) Cohabitation constraint:

$$\forall\ (M_i , M_q) \in SC,\ \forall w \in CS\text{: } e_{iq} \notin w, \text{ where } i,\ q \in \{1, ..., m\}.$$

(3) Non-cohabitation constraint:

$$\forall\ (M_i, M_q) \in SN,\ \exists\ w \in CS\text{: } e_{iq} \in w \text{, where } \ i,q \in \{1, ....., m\}.$$

*2.2.6 Objective function*

Let $S$ be the set of cut subsets that respect the previous constraints. The problem is to find a solution $CS^* \in S$, such that:

$$Z(CS^*) = \min_{CS \in S} T(CS)$$

This means to seek a cut subset that respects all the constraints and has the minimum amount of intercellular traffic.

***2.3 Theoretic preliminaries to the cut-based approach***

Since a solution is a graph partition, it can, *in the case of a connected graph instance*, be represented by a sum, using the boolean operator *OR* , denoted ⊕, of cuts (A cut is a subset of edges that can be associated with a subset of vertices $A$ for which all these edges have exclusively and exactly one endpoint in $A$). The solution of figure 1, for example, can be defined by the sum of two of the depicted cuts $w_1,w_2,w_3$ .

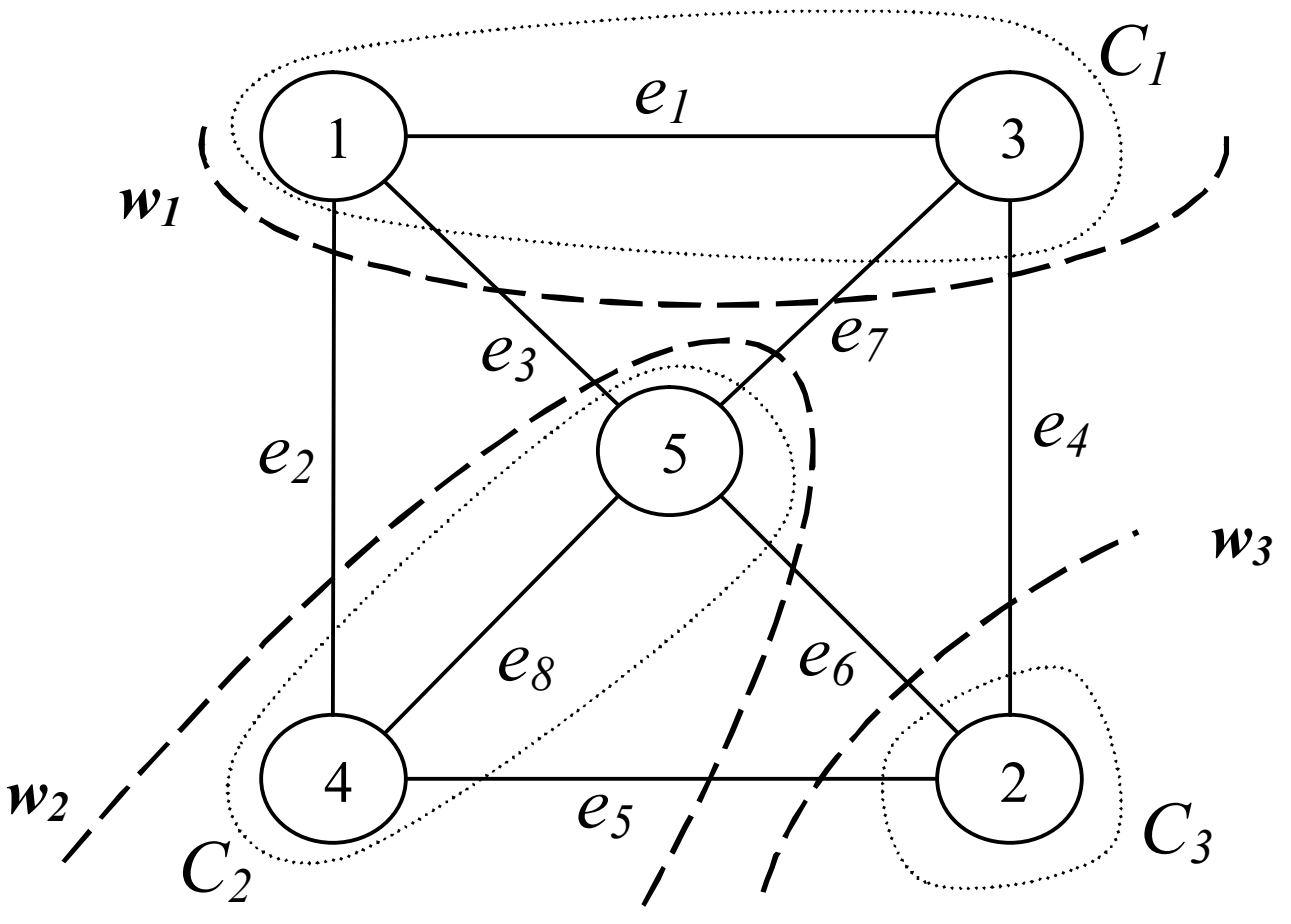

**Fig. 1** A solution instance with its constructor cuts.

For instance, the sum of $w_1$=(0,1,1,1,0,0,1,0) and $w_2$=(0,1,1,0,1,1,1,0) yields $w_1 \oplus w_2$ =(0,1,1,1,1,1,1,0), which is sufficient to determine the associated solution. In fact, the cuts are represented by binary vectors in which the ones indicate the associated edges. For example, $w_1$=(0,1,1,1,0,0,1,0) is constructed by $e_2,e_3,e_4,e_7$ because their corresponding values equal one.

The interpretation of the obtained solution vector is slightly different in the fact that ones correspond to *inter*cellular edges and zeros to *intra*cellular edges. According to this interpretation, the obtained sum (0,1,1,1,1,1,1,0) sets $e_1$ and $e_8$ intracellular and the remainder edges intercellular, yielding the solution $C$={{$M_1$,$M_3$},{$M_2$},{$M_4$,$M_5$}} of figure 1.

A partition being a sum of cuts yields to the fact that the search space can be covered using these "graph creatures". However, the edge-based binary codification is not directly suitable because a random binary vector is not necessarily a cut. Fortunately, we can overcome this first hurdle using cut properties in graph theory. In fact, cuts define a vector space that can be covered by the XOR operator and a subset of only $m$-1 cuts, the base of the vector space. There are several manners to get a cut base. Among the simplest ways, the direct method consists of choosing any $m$-1 vertices and then getting the cuts associated with the singletons of the chosen vertices. For instance, for the graph of figure 1, if we choose the singletons {$M_1$},{$M_2$},{$M_3$},{$M_4$} their associated cuts, i.e. $w$({$M_1$})=(1,1,1,0,0,0,0,0), $w$({$M_2$})=(0,0,0,1,1,1,0,0), $w$({$M_3$})=(1,0,0,1,0,0,1,0) and $w$({$M_4$})=(0,1,0,0,1,0,0,1), define a cut base. Any cut has a unique representation as a XOR sum of basic cuts. For example, $w_1$= $w$({$M_1$}) XOR $w$({$M_3$}) and $w_2$= $w$({$M_1$}) XOR $w$({$M_2$}) XOR $w$({$M_3$}). For further theoretic issues on cut properties see (Picard 1972).

## 3. The cut-based GA

Genetic algorithms are one of the famous optimization approaches that mimic natural processes. In this section, the general principles of GA are first presented (Boulif 2006), followed by a description of the GA applied to the MCF problem.

### 3.1 *Principles of the genetic algorithm approach*

Due to his publication, *« Adaptation in Natural and Artificial Systems »*, J.Holland (Holland 1975) is considered to be the founder of the modern Genetic Algorithms. These algorithms are based on an analogy to the biological phenomenon of natural selection. First, a chromosome structure is set to represent the solutions of the problem. Afterwards, an initial solution population of chromosomes is generated, either randomly or using a given heuristic. Then, members of the population are selected, based on an evaluation function, called fitness. The fitness associates a value to each member according to its objective function. The higher a member's fitness value, the more likely it is to be selected. Thus, the less fit individuals are replaced by those who perform better. Genetic operators are then applied to the selected members to produce a new population generation. This process is repeated until achieving a certain stopping criterion.

Implementing genetic algorithms requires that the following aspects be defined:

- the structure of the genetic code used for representing solutions;
- the method for generating the initial population;
- an adaptation function for evaluating the fitness of each member of the population;
- the genetic operators used for producing a new generation; and
- certain control parameter values (eg. population size, number of iterations, genetic operator probabilities).

### 3.2 *GA implementation*

#### *3.2.1 Cut-based encoding*

The graph theory model (see section 2) allows solutions to be encoded in a chain of $K\times(m-1)$ binary alleles, where $K=\left[\frac{m}{N}\right]$ ($[x]$ denotes the integer part of $x$). In other words, this chain is composed of $K$ parts of length $m$-1. Each part allows defining a cut by specifying the basic cuts that construct it with the XOR sum. Thus each one of the $K$ parts yields a cut. Afterwards, combined with an OR sum, these cuts define the associated partition solution. The definition of $K$ is due to the fact that a typical good solution has probably a moderate number of cells which cannot be less than $\frac{m}{N}$ and thus $K$ cuts are sufficient to construct a good solution. By supposing $K$ to be equal to 3, the three-cell solution of figure 1 would be coded by the following chain:

| *part 1* | | | | *part 2* | | | | *part 3* | | | |
|---|---|---|---|---|---|---|---|---|---|---|---|
| $w(\{M_1\})$ | $w(\{M_2\})$ | $w(\{M_3\})$ | $w(\{M_4\})$ | $w(\{M_1\})$ | $w(\{M_2\})$ | $w(\{M_3\})$ | $w(\{M_4\})$ | $w(\{M_1\})$ | $w(\{M_2\})$ | $w(\{M_3\})$ | $w(\{M_4\})$ |
| 1 | 0 | 1 | 0 | 1 | 1 | 1 | 0 | 0 | 0 | 0 | 0 |

The interpretation of this chromosome structure is straightforward: part 1 uses $w(\{M_1\})$ and $w(\{M_3\})$ to define the first cut $w_1$. The second uses $w(\{M_1\})$, $w(\{M_2\})$ and $w(\{M_3\})$ yielding $w_2$. In the third part all the alleles are equal to zero, and thus no cut is generated. The two cuts, $w_1$ and $w_2$, yield by an OR sum the associated partition.

The most important advantage of binary coding is that GAs are positively sensitive to reduced alphabets (0 and 1 in this case). With a binary alphabet, it becomes easier to the GA to detect the good blocks of the individuals' codes. However, the cut-based GA suffers from a high level of redundancy. In fact, a solution is not affected by swapping its parts and, furthermore, we can have two equal parts. To overcome this second hurdle, we sort without repetition each solution from the right to the left part. This sorting considers for each part the decimal number taken from the sub-string of the part when it is supposed binary coded (see the appendix). For example, the previous chain is coded in decimal as follows:

| *part 1* | *part 2* | *part 3* |
|---|---|---|
| 10 | 14 | 0 |

Thus, part 2 sub-string must be put first, then part1 and finally part3. After this *sorting procedure*, there will be no equal parts except for a possible sequence of zeros tail.

*3.2.2 Initial population*

The initial population is randomly generated without repetition. To get a solution, we generate $K$ integers from the interval $[0, 2^{m-1}-1]$ (see the appendix). Each integer, when it is converted to binary code, will give a part of the solution chain. After applying the sorting procedure, if there is no equivalent member in the population, the solution is accepted.

*3.2.3 Fitness and selection*

To allow the GA to get advantages of the good information unfeasible solutions can hide, the fitness is calculated in such a manner that enables those solutions to contribute to the exploration scheme (Boulif 2006).

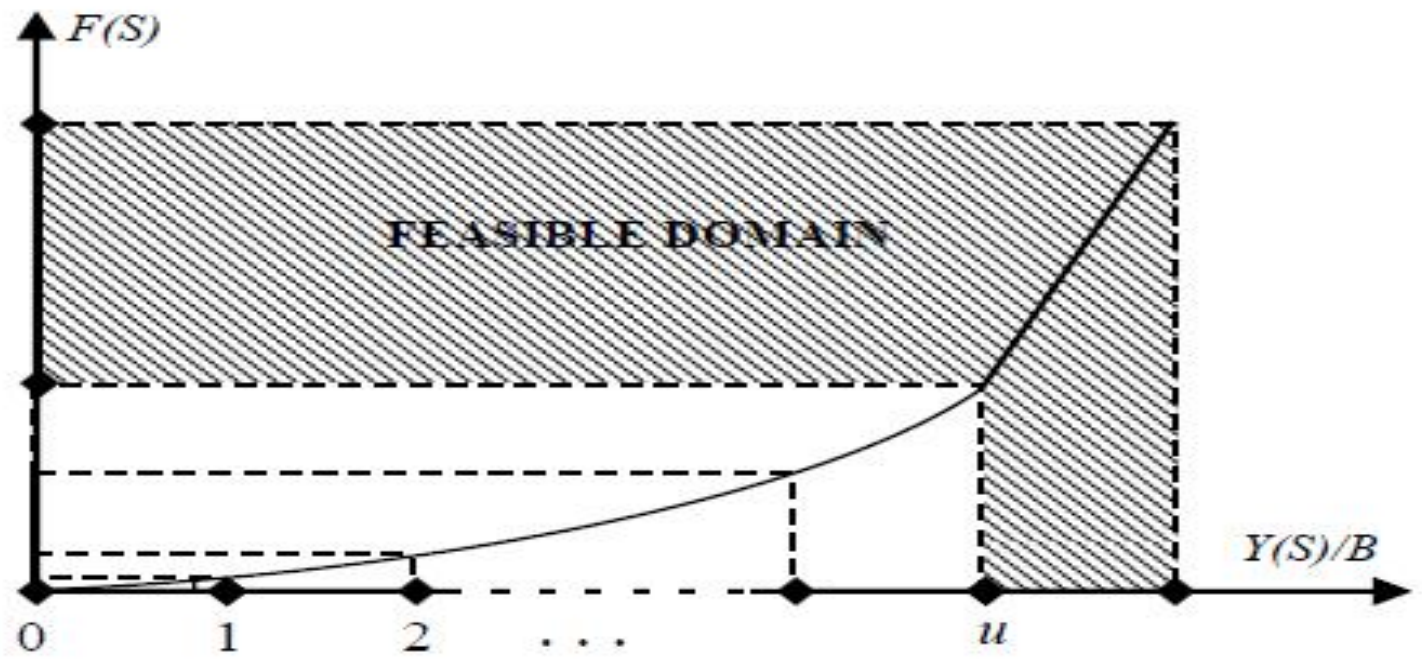

**Fig. 2** Fitness fine-tuning procedure (Boulif 2006)

First, the minimization problem is transformed into a maximization problem via the formula, $Z'(S) = B - Z(S)$, where $Z(S)$ is the value of the objective function for a given solution $S$, and $B$ is an upper bound of $Z(S)$. Second, the obtained value is translated using the formula, $Y(S)=Z'(S)+(u-v(S))\times B$, where $u$ is the number of constraints, and $v(S)$ is the number of $S$ unverified constraints. The $Y(S)$ value obtained can then be fine-tuned using a function that allows the feasible domain to be widened (see figure 2).

*Remark* Broadly speaking, $B$ can be set to any upper bound of the objective function to be minimized. However, it is desirable to choose the most tightened value. In our problem $B$ is set to the sum of the overall product flows.

The "Roulette wheel" random procedure was used to select an individual (Goldberg 1989). On this wheel, each individual in the population has a slot proportional to its fitness. Spinning the wheel as many as required defines the individuals eligible to the crossover.

*3.2.4 Crossover and mutation*

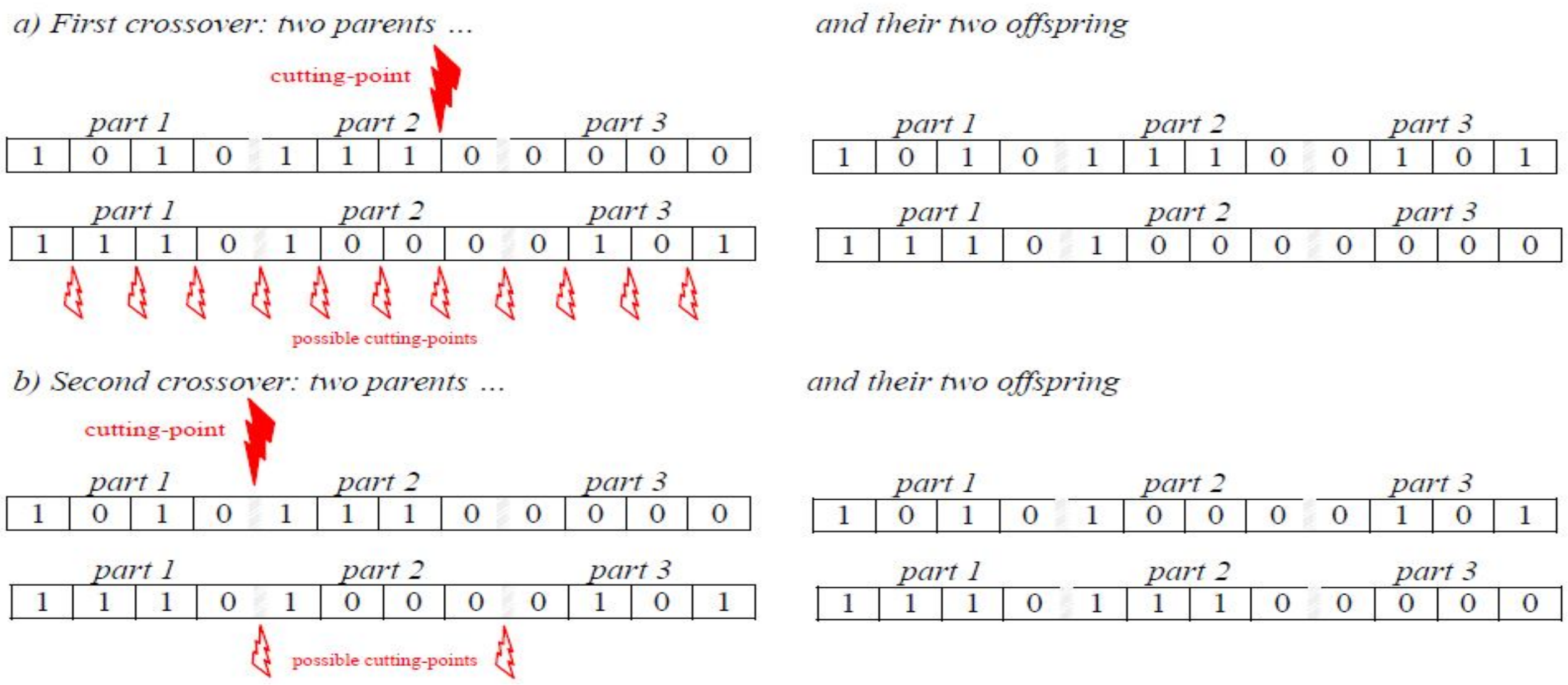

**Fig. 3.** Crossovers

For simplicity, we have opted for one-cutting-point crossovers. The first one is classical and allows putting the cutting point in any random point of the chain. The second crossover allows putting the cutting point only between consecutive cut parts (see figure 3). The ratio of individuals that will undergo a crossover operator is defined by the parameter *Pc*. The rest give up their places to other randomly generated members. The mutation operator consists of randomly choosing a ratio of *Pm* members. For each one, a cut part is replaced by another one randomly generated. For the parameter settings, empirical experimentation has been conducted to choose the parameter values that push the GA to perform at its best in a small amount of time (less than one minute).

*3.2.5 Stopping criterion*

After the selection-crossover-mutation process, the sorting procedure is applied on each individual of the population. The best individual that has been saved before the three step process is then reinserted in the population (elitism). This process is repeated until a certain number of iterations $i_{max}$ is reached.

*3.2.6 Cut based GA algorithm*

The pseudo code of the GA we implemented is as follows:

1. Get random population (with feasible or unfeasible individuals without repetition);
   [Apply the sorting procedure on each individual of the population;]
2. Evaluate population using fine tuning procedure (see 3.2.3);
3. Repeat
   - Save the best fitted individual;
   - Get mating population : Select *Pc* proportion of individuals from population using the Roulette Wheel Selection procedure;
   - Apply crossover (pick one of the two randomly) on the mating population and replace parents by their offspring; Generate the rest (1-*Pc*) of the population randomly;
   - Apply mutation on the resulting population with *Pm* ratio;
   - Reinsert the best fitted individual;
   - [Apply the sorting procedure on each individual of the population;]
   - Re-evaluate population;

   Until $i_{max}$ iterations;

## 4. Computational results

Two variants of the cut based genetic algorithm were implemented. The first one does not implement the sorting procedure whereas the second does. The two cut based GA were also compared with the edge based GA (Boulif 2006) and our implementation of the well-known Kmeans clustering (Lloyd 1982). Our implementation of Kmeans initializes the centroids by using random rows of the machine-to-machine-product-flow matrix. In addition, since Kmeans uses a known number of clusters (cells), we calls Kmeans for all the integer values in [m/N, m-1] interval; hence, we call it multiKmeans. The corresponding pseudo code is as follows:

Loop for k=[m/N] to m-1 do

- Call kMeans(input: number of cell (clusters) k, machine-to-machine-product-flow adjacency matrix, output: machine partition solution);
- update the best solution if outperformed;

Loop end;

The four applications were processed on a Core *i*3 microcomputer with a clock speed of 2,1 GHz and 3.8 Go of RAM managed by a 32-bit Linux operating system. We coded them by using a C++ compiler. In the following paragraphs, the four methods are referred to as **CGA** for the **C**ut based **G**enetic **A**lgorithm without the sorting procedure, **SCGA** for the **C**ut based **G**enetic **A**lgorithm with the **S**orting procedure, **EGA** for the **E**dge based **G**enetic **A**lgorithm, and **multiKmeans**.

Four examples are taken from the literature (Boulif 2006) and a fifth example has been generated randomly. They are sorted according to their size, assumed to be equal to the product p×m (number of products × number of machines).

The five examples have a size of 20×8, 20×20, 40×20, 51×20 and 100×50 respectively; and the maximum number of machines per cell is set to 5 for all but the third example where it is set to 4 and the fifth where it is set to 7 then 15.

Aiming at using moderate resources, we considered the evolutionary methods with the following parameter values:

- Population size : 100, 200, 300, 400 and 500;
- Number of generations : 100, 200 and 300;

For the crossover and mutation rates, values in the interval [0.6,0.8] for the first, and in [0.01,0.05] for the second have shown comparable performances. Therefore, we have chosen to set them to 0.7 and 0.03 respectively in the presented results.

We run each one of the three methods for 20 times, and then we have reported the best average traffic and the best solution with its own computational running time. The obtained results are reported in Table 1.

| Parameters | | First example (size=20×8) | | | | | | | | | Third example (size=40×25) | | | | | | | | | fifth example (size=100×35, N=7) | | | | | | | | |
|---|---|---|---|---|---|---|---|---|---|---|---|---|---|---|---|---|---|---|---|---|---|---|---|---|---|---|---|---|
| | | EGA | | | CGA | | | SCGA | | | EGA | | | CGA | | | SCGA | | | EGA | | | CGA | | | SCGA | | |
| Pop. size | Gen. # | Avg. traffic | Best traffic | Cpu (s) | Avg. traffic | Best traffic | Cpu (s) | Avg. traffic | Best traffic | Cpu (s) | Avg. traffic | Best traffic | Cpu (s) | Avg. traffic | Best traffic | Cpu (s) | Avg. traffic | Best traffic | Cpu (s) | Avg. traffic | Best traffic | Cpu (s) | Avg. traffic | Best traffic | Cpu (s) | Avg. traffic | Best traffic | Cpu (s) |
| 100 | 100 | 16.0 | 13 | 0.07 | 26.0 | 20 | 0.13 | 27 | 22 | 0.15 | - | UF* | - | 85.0 | 80 | 0.25 | 83.7 | 75 | 0.24 | - | UF | - | 515.1 | 511 | 3.00 | 487.8 | 453 | 3.42 |
| | 200 | 19.8 | 17 | 0.13 | 27.6 | 22 | 0.25 | 24.5 | 21 | 0.29 | - | UF | - | 85.0 | 81 | 0.48 | 80.5 | 70 | 0.47 | - | UF | - | 516.1 | 513 | 6.02 | 486.9 | 435 | 6.89 |
| | 300 | 25.2 | 19 | 0.19 | 27.4 | 24 | 0.40 | 25.0 | 20 | 0.44 | 78.1 | 66 | 0.16 | 84.9 | 81 | 0.73 | 80.1 | 72 | 0.72 | - | UF | - | 513.7 | 503 | 8.84 | 477.7 | 429 | 9.93 |
| 200 | 100 | 15.5 | 13 | 0.13 | 28.7 | 21 | 0.38 | 26.5 | 22 | 0.47 | 63.1 | 59 | 0.12 | 84.6 | 81 | 0.47 | 77.3 | 73 | 0.49 | - | UF | - | 509.2 | 469 | 6.00 | 476.6 | 446 | 6.96 |
| | 200 | 16.4 | 13 | 0.27 | 27.6 | 20 | 0.95 | 25.7 | 17 | 0.88 | 74.7 | 58 | 0.22 | 82.8 | 79 | 0.97 | 75.8 | 62 | 1.01 | 531.2 | 497 | 0.78 | 496.9 | 447 | 12.21 | 469.4 | 451 | 13.61 |
| | 300 | 19.2 | 13 | 0.41 | 27.6 | 19 | 1.36 | 26.1 | 19 | 1.25 | 67.5 | 49 | 0.34 | 83.5 | 81 | 1.46 | 74.6 | 68 | 1.51 | - | UF | - | 498.3 | 441 | 17.69 | 455.1 | 423 | 20.71 |
| 300 | 100 | 18.6 | 15 | 0.22 | 28.6 | 21 | 0.68 | 24.1 | 17 | 0.77 | 60.0 | 51 | 0.19 | 82.7 | 79 | 0.78 | 74.8 | 68 | 0.82 | - | UF | - | 503.0 | 436 | 9.94 | 477.5 | 447 | 10.34 |
| | 200 | 20.0 | 13 | 0.42 | 27.2 | 20 | 1.59 | 26.0 | 20 | 1.83 | 71.8 | 56 | 0.35 | 82.3 | 80 | 1.55 | 73.6 | 64 | 1.61 | 529.6 | 465 | 1.49 | 502.2 | 453 | 18.48 | 463.4 | 435 | 20.37 |
| | 300 | 19.0 | 13 | 0.65 | 26.0 | 21 | 2.58 | 27.7 | 22 | 2.56 | 67.2 | 47 | 0.52 | 82.3 | 80 | 2.36 | 73.4 | 61 | 2.68 | - | UF | - | 484.7 | 422 | 25.19 | 465.1 | 434 | 31.61 |
| 400 | 100 | 17.4 | 13 | 0.32 | 24.6 | 20 | 1.26 | 25.2 | 20 | 1.53 | 65.7 | 54 | 0.25 | 79.1 | 60 | 1.13 | 76.8 | 62 | 1.25 | 527.7 | 476 | 1.14 | 490.2 | 436 | 12.54 | 479.9 | 441 | 14.07 |
| | 200 | 18.9 | 15 | 0.63 | 27.1 | 25 | 3.04 | 26.6 | 22 | 2.65 | 61.8 | 53 | 0.50 | 81.2 | 77 | 2.23 | 71.5 | 65 | 2.35 | 529.6 | 495 | 1.97 | 491.5 | 432 | 24.85 | 458.6 | 421 | 27.65 |
| | 300 | 18.2 | 13 | 0.93 | 26.6 | 22 | 3.96 | 27.4 | 22 | 4.22 | 67.2 | 58 | 0.74 | 81.0 | 69 | 3.53 | 71.0 | 65 | 3.74 | 526.2 | 463 | 2.37 | 479.5 | 429 | 32.90 | 461.8 | 413 | 39.99 |
| 500 | 100 | 17.3 | 13 | 0.40 | 27.4 | 19 | 1.67 | 25.9 | 22 | 2.57 | 62.1 | 56 | 0.40 | 80.5 | 62 | 1.51 | 75.6 | 68 | 1.60 | 530.4 | 482 | 1.36 | 492.2 | 447 | 15.50 | 466.6 | 432 | 18.12 |
| | 200 | 16.7 | 13 | 0.82 | 25.4 | 19 | 4.34 | 26.3 | 19 | 3.99 | 64.9 | 50 | 0.62 | 80.3 | 72 | 2.98 | 73.1 | 57 | 3.37 | 528.9 | 492 | 2.95 | 483.0 | 433 | 31.94 | 459.4 | 435 | 36.26 |
| | 300 | 18.45 | 13 | 1.21 | 26.0 | 19 | 6.10 | 26.6 | 19 | 6.56 | 62.0 | 49 | 0.93 | 76.7 | 65 | 4.54 | 70.2 | 61 | 4.98 | 520.3 | 455 | 3.19 | 464.9 | 422 | 45.02 | 443.1 | 418 | 54.56 |

| Parameters | | Second example (size=20×20) | | | | | | | | | Fourth example (size=51×20) | | | | | | | | | sixth example (size=100×50, N=15) | | | | | | | | |
|---|---|---|---|---|---|---|---|---|---|---|---|---|---|---|---|---|---|---|---|---|---|---|---|---|---|---|---|---|
| | | EGA | | | CGA | | | SCGA | | | EGA | | | CGA | | | SCGA | | | EGA | | | CGA | | | SCGA | | |
| Pop. size | Gen. # | Avg. traffic | Best traffic | Cpu (s) | Avg. traffic | Best traffic | Cpu (s) | Avg. traffic | Best traffic | Cpu (s) | Avg. traffic | Best traffic | Cpu (s) | Avg. traffic | Best traffic | Cpu (s) | Avg. traffic | Best traffic | Cpu (s) | Avg. traffic | Best traffic | Cpu (s) | Avg. traffic | Best traffic | Cpu (s) | Avg. traffic | Best traffic | Cpu (s) |
| 100 | 100 | 30.1 | 27 | 0.06 | 45.9 | 43 | 0.13 | 38.9 | 38 | 0.11 | 211.2 | 184 | 0.09 | 192.3 | 166 | 0.44 | 183.7 | 167 | 0.38 | 506.7 | 398 | 0.18 | 461.5 | 449 | 0.71 | 390.0 | 350 | 0.72 |
| | 200 | 30.4 | 25 | 0.10 | 44.0 | 41 | 0.29 | 38.4 | 33 | 0.24 | 185.7 | 165 | 0.18 | 187.7 | 153 | 0.93 | 183.5 | 139 | 0.85 | - | UF | - | 457.3 | 446 | 1.41 | 378.5 | 342 | 1.48 |
| | 300 | 27.5 | 22 | 0.15 | 41.8 | 38 | 0.49 | 37.0 | 30 | 0.34 | 199.5 | 168 | 0.26 | 190.6 | 167 | 1.32 | 180.8 | 145 | 1.25 | - | UF | - | 449.0 | 392 | 2.27 | 369.9 | 339 | 2.21 |
| 200 | 100 | 27.8 | 24 | 0.10 | 43.9 | 36 | 0.34 | 39.7 | 37 | 0.28 | 197.4 | 186 | 0.18 | 184.0 | 152 | 0.95 | 176.3 | 152 | 0.90 | 527.4 | 493 | 0.36 | 448.5 | 436 | 1.46 | 373.4 | 345 | 1.58 |
| | 200 | 28.4 | 24 | 0.21 | 42.7 | 39 | 0.62 | 36.9 | 29 | 0.59 | 187.8 | 132 | 0.37 | 173.8 | 154 | 1.93 | 173.6 | 142 | 1.79 | - | UF | - | 447.2 | 424 | 2.96 | 359.5 | 340 | 3.12 |
| | 300 | 29.5 | 26 | 0.31 | 41.5 | 37 | 0.96 | 35.9 | 31 | 0.86 | 200.9 | 153 | 0.54 | 172.8 | 158 | 3.25 | 167.8 | 140 | 2.81 | - | UF | - | 431.1 | 363 | 4.94 | 367.7 | 344 | 4.96 |
| 300 | 100 | 27.7 | 24 | 0.17 | 40.9 | 38 | 0.53 | 36.2 | 32 | 0.52 | 202.7 | 162 | 0.29 | 173.0 | 146 | 2.17 | 169.3 | 139 | 1.81 | - | UF | - | 448.5 | 439 | 2.24 | 368.9 | 350 | 2.41 |
| | 200 | 28.4 | 25 | 0.34 | 41.6 | 39 | 1.35 | 35.5 | 26 | 1.08 | 199.2 | 178 | 0.57 | 163.7 | 144 | 4.04 | 165.1 | 132 | 3.18 | 525.8 | 390 | 1.77 | 434.7 | 363 | 4.51 | 367.9 | 341 | 5.09 |
| | 300 | 26.0 | 22 | 0.54 | 40.8 | 36 | 1.62 | 34.8 | 31 | 1.55 | 183.5 | 102 | 0.86 | 164.1 | 145 | 7.3 | 165.3 | 143 | 5.00 | 522.2 | 415 | 2.40 | 432.0 | 390 | 7.91 | 363.4 | 326 | 7.79 |
| 400 | 100 | 28.1 | 24 | 0.23 | 41.2 | 38 | 0.79 | .34.9 | 30 | 0.80 | 198.7 | 164 | 0.4 | 171.7 | 148 | 3.3 | 158.7 | 136 | 2.34 | 510.7 | 394 | 0.90 | 443.1 | 417 | 3.11 | 364.9 | 334 | 3.27 |
| | 200 | 25.6 | 19 | 0.45 | 40.3 | 36 | 1.52 | 33.3 | 28 | 1.72 | 183.4 | 143 | 0.83 | 162.0 | 143 | 6.75 | 156.3 | 120 | 4.93 | 519.3 | 438 | 1.61 | 427.9 | 352 | 7.38 | 363.7 | 331 | 7.51 |
| | 300 | 25.9 | 19 | 0.73 | 39.3 | 35 | 2.69 | 33.2 | 30 | 2.45 | 187.1 | 113 | 1.23 | 162.0 | 141 | 10.17 | 161.1 | 129 | 8.75 | 514.4 | 403 | 2.59 | 413.4 | 336 | 10.40 | 360.8 | 339 | 13.75 |
| 500 | 100 | 26.9 | 22 | 0.30 | 39.5 | 37 | 1.07 | 34.8 | 31 | 1.04 | 183.8 | 118 | 0.53 | 170.5 | 152 | 5.36 | 161.7 | 140 | 3.78 | 526.1 | 460 | 1.03 | 435.4 | 336 | 4.14 | 364.6 | 338 | 4.40 |
| | 200 | 26.0 | 20 | 0.60 | 39.7 | 37 | 2.33 | 31.6 | 25 | 2.18 | 196.4 | 146 | 1.06 | 161.2 | 140 | 10.53 | 155.5 | 136 | 9.04 | 523.8 | 349 | 2.15 | 428.6 | 369 | 8.18 | 366.0 | 341 | 9.61 |
| | 300 | 24.1 | 21 | 0.90 | 38.7 | 34 | 3.40 | 33.2 | 25 | 3.19 | 192.7 | 155 | 1.74 | 160.0 | 147 | 16.11 | 150.3 | 110 | 11.52 | 486.2 | 394 | 3.78 | 405.0 | 345 | 16.16 | 358.5 | 331 | 14.83 |

* UF : unfeasible solution.

**Table 1.** Computational results

In the other hand, multiKmeans gave 32, 50, 194, UF, UF and 416 in less than 0.1 second.

As it can be deduced, the three evolutionary methods are far better than Kmeans based approach that does implement neither a constraint handling routine nor a mechanism to avoid local optima trapping.

From Table1, in the first example, EGA when compared to the other methods had a twofold performance: it reached a better value of the objective function in a lesser running time. In the second example, EGA was still able to reach a better value of the traffic. However, the average traffic of EGA reveals a great difficulty in reaching these best values. Indeed, CGA and SCGA were clearly more responsive to increasing the population size by reaching far better solution in the average, as depicted in figure 4 for the fourth example.

Figure 4 Avergae traffic comparison for example 4.

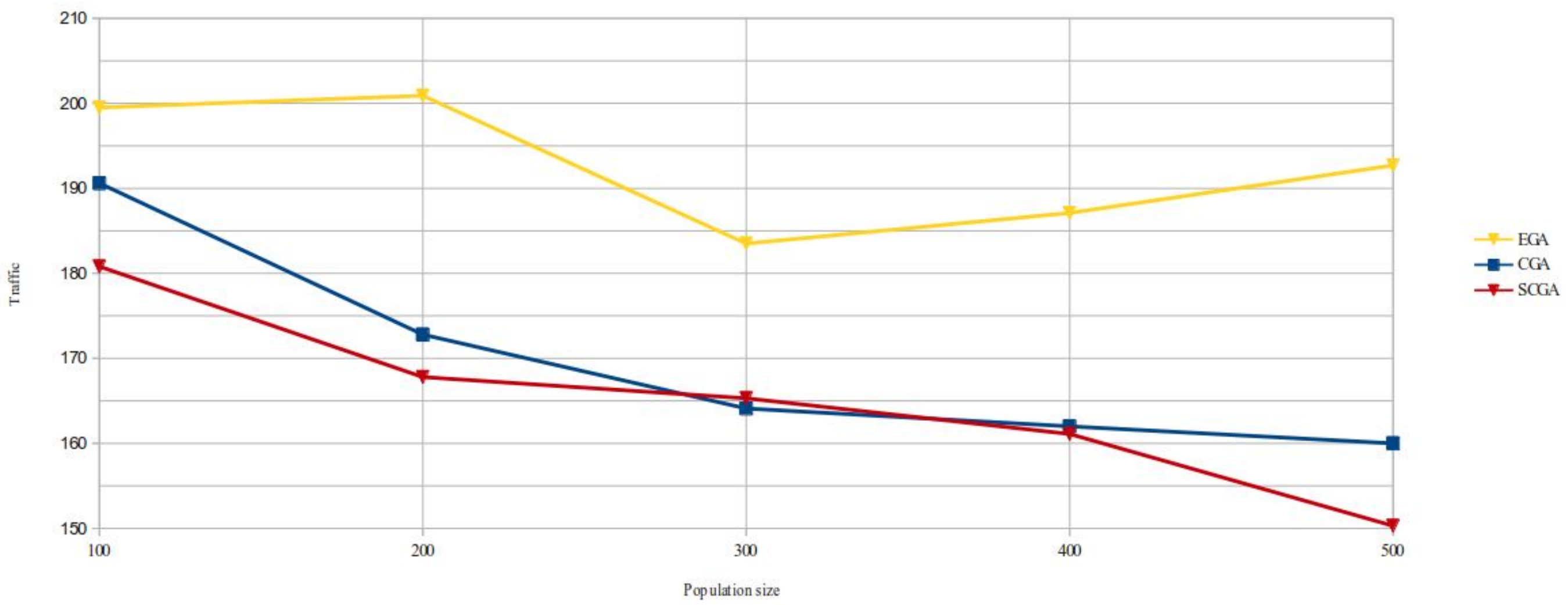

For the largest example, we have considered two values for the maximum cell size; that is, N=7 and N=15. In both variants, EGA struggles to reach feasible solutions in small population sizes and limited number of generations in comparison to its counterparts. However, SCGA that gave the best performances in average requires more time resources especially when the expected number of cells (clusters) for good solutions grows. Indeed, when the maximum cell size decreases, cut based methods need more graph cuts to construct feasible solutions because with the overhead of the sorting procedure SCGA needs more time to achieve its optimisation process.

## 5. Conclusions

This paper deals with cell formation witch is one of the main problems to be solved when dealing with cellular manufacturing. A genetic algorithm with a new graph-cut-based encoding representation is proposed. The performance of the new GA is tested on a set of numerical examples and compared with other methods. The cut based GA has proven to be more able to reach feasible areas with low resources especially when the expected number of cells for good solutions is moderate.

We suggest continuing this work in the following directions. First, we are interested by adopting other ways for constructing the cut base. Inspecting then their influence on the performance of the cut based GA will be a good path of investigation. Second, the compared evolutionary methods are from the same family as they belong to the edge-based approach. This suggests a co-evolutionary solving approach is very promising. Finally, the branch and bound enhancement (Boulif 2006) being closer to the cut based GA than the edge based one, it seems that a hybridization of the two methods is another promising research path.

## References

Boulif M., Atif K., A new branch-&-bound-enhanced GA for the manufacturing cell formation problem. *Computers & Operations research*, 2006, 33:2219-2245.

Darla SP., Naiju CD., Sagar PV. and Likhit BV., Optimization of Inter Cellular Movement of Parts in Cellular Manufacturing System Using Genetic Algorithm, 2014, *Research Journal of Applied Sciences, Engineering and Technology*, 7, 1, 165-168.

Goldberg DE., Genetic Algorithms in Search, Optimization and Machine Learning, *Addison Wesley, Reading, MA*. 1989.

Gupta T., and Seifoddini H., Production data based similarity coefficient for machine-component grouping decision in the design of a cellular manufacturing system., **28:**1247-69. 1990.

Gupta YP., Gupta MC., Kumar A., Sundram C. Minimizing total intercell and intracell moves in cellular manufacturing: a genetic algorithm approach. *Int. J. Integ. Manuf.*, 1995, 8, 2 92-101.

Holland JH., Adaptation in natural and artificial systems. Ann Arbor: University of Michigan Press. 1975.

Kaparthi S., Suresh N. C., Machine-component cell formation in group technology: a neural network approach, *Int. J. Prod. Res.,* 1992, 30, 6 1353-1367.

Lloyd SP., Least squares quantization in PCM, *IEEE Transactions on Information Theory*, 1982, 28 (2): 129–137.

Logendran R., Ramakrishna P., Manufacturing cell formation in the presence of Lot splitting and multiple units of same machine, *Int. J. Prod. Res.,* 1995, 33, 675-693.

Mahapatra SS., Pandian RS., Genetic cell formation using ratio level data in cellular manufacturing systems, *Int J Adv Manuf Technol.*, 2008, 38:630–640.

McAULEY J., Machine Grouping For Efficient Production, *Production Engineer,* 1972, 52(2): 53-57.

Merchichi S., Boulif M., Constraint-driven exact algorithm for the manufacturing cell formation problem. *European Journal of Industrial Engineering*, 2015, 9(6):717-743.

Nunkaew W., Phruksaphanrat B., Effective fuzzy multi–objective model based on perfect grouping for manufacturing cell formation with setup cost constrained of machine duplication, *Songklanakarin J. Sci. Technol.*, 2013, 35(6) : 715-726.

Picard CF., Graphes et questionnaires: Tome1 Graphes, Gauthier-Villars. 1972.

Seifoddini H., Single linkage vs average linkage clustering in machine cells formation applications, *Computers & Industrial Engineering,* 1989, 16, 419-426.

Souilah A., Les systèmes cellulaires de production : Agencement intracellulaire. PhD dissertaion*,* University of Metz. 1994.

Venugopal V., Narendran TT., Cell formation in manufacturing systems through simulated annealing: an experimental evaluation, *Eur J. Oper. Res.,* 1992, 63, 409-422.

Yin Y., Yasuda K., Similarity coefficient methods applied to the cell formation problem: A taxonomy and review, *International Journal of Production Economics*, 2006, 101, 2, 329–352.

## APPENDIX A

### A.1 The number of cuts

In a connected graph with $m$ vertices there are $2^{m-1}$-1 cuts. Indeed, a cut splits the graph in two complementary subsets of vertices. Since we can get $2^m$ subsets from a set of cardinality $m$, we have $2^{m-1}$ couples of complementary subsets. By taking away the special couple of the complete and the empty sets yields the number of cuts.

### A.2 Representing cuts in decimal

We consider the graph of figure 1. The number of cuts is $2^{5-1}$-1=15. Each cut can be derived from a combination of basic cuts that yields a unique integer number. This integer representation is very useful since it defines a bijection between the set of cuts and the set of integers in the interval [1, $2^{m-1}$-1].

**Table A.1** Different representations of cuts

| Integer representation | Representation with basic cuts | | | | Edge representation |
|---|---|---|---|---|---|
| | $w(\{M_1\})$ | $w(\{M_2\})$ | $w(\{M_3\})$ | $w(\{M_4\})$ | $(e_1,e_2,e_3,e_4,e_5,e_6,e_7,e_8)$ |
| **1** | **1** | **0** | **0** | **0** | **(1,0,1,0,0,0,0,1)** |
| **2** | **0** | **1** | **0** | **0** | **(0,0,0,1,1,1,0,0)** |
| 3 | 1 | 1 | 0 | 0 | (1,0,1,1,1,1,0,1) |
| **4** | **0** | **0** | **1** | **0** | **(1,0,0,1,0,0,1,0)** |
| 5 | 1 | 0 | 1 | 0 | (0,0,1,1,0,0,1,1) |
| 6 | 0 | 1 | 1 | 0 | (1,0,0,0,1,1,1,0) |
| 7 | 1 | 1 | 1 | 0 | (0,0,1,0,1,1,1,1) |
| **8** | **0** | **0** | **0** | **1** | **(0,1,0,0,1,0,0,1)** |
| 9 | 1 | 0 | 0 | 1 | (1,1,1,0,1,0,0,0) |
| 10 | 0 | 1 | 0 | 1 | (0,1,0,1,0,1,0,1) |
| 11 | 1 | 1 | 0 | 1 | (1,1,1,1,0,1,0,0) |
| 12 | 0 | 0 | 1 | 1 | (1,1,0,1,1,0,1,1) |
| 13 | 1 | 0 | 1 | 1 | (0,1,1,1,1,0,1,0) |
| 14 | 0 | 1 | 1 | 1 | (1,1,0,0,0,1,1,1) |
| 15 | 1 | 1 | 1 | 1 | (0,1,1,0,0,1,1,0) |

Table A.1 depicts the set of cuts with their different representations for the previous graph. Note that basic cuts (bold typed) are powers of two.

**APPENDIX B**

The operation sequences for the randomly generated example (fifth example in the computational results) are depicted in Table B.1.

**Table B.1** Operation sequences for example 5

| | M 1 | M 2 | M 3 | M 4 | M 5 | M 6 | M 7 | M 8 | M 9 | M 10 | M 11 | M 12 | M 13 | M 14 | M 15 | M 16 | M 17 | M 18 | M 19 | M 20 | M 21 | M 22 | M 23 | M 24 | M 25 | M 26 | M 27 | M 28 | M 29 | M 30 | M 31 | M 32 | M 33 | M 34 | M 35 | M 36 | M 37 | M 38 | M 39 | M 40 | M 41 | M 42 | M 43 | M 44 | M 45 | M 46 | M 47 | M 48 | M 49 | M 50 |
|---|---|---|---|---|---|---|---|---|---|---|---|---|---|---|---|---|---|---|---|---|---|---|---|---|---|---|---|---|---|---|---|---|---|---|---|---|---|---|---|---|---|---|---|---|---|---|---|---|---|---|
| P1 | · | · | · | · | · | · | · | · | · | · | · | · | · | · | · | · | · | · | · | · | · | · | · | · | · | · | · | · | · | · | · | · | · | · | · | · | · | · | · | · | · | · | 1, | · | · | · | · | · | · | 2, |
| P2 | · | · | · | · | · | · | 4, | · | · | · | 2, | · | · | · | · | · | 1, | · | · | · | · | · | · | · | · | · | · | · | · | · | · | · | · | · | · | · | · | · | · | 3, | · | · | · | · | 5, | · | · | · | · | · |
| P3 | · | · | · | · | · | · | · | · | · | · | · | · | · | · | · | · | · | · | · | · | · | · | · | · | · | · | 1, | · | · | 2, | · | · | · | · | · | · | · | · | · | · | · | · | · | · | · | · | · | · | · | · |
| P4 | · | · | · | · | · | · | · | · | · | · | · | · | · | · | · | 4, | · | 3, | · | · | · | · | 1, | · | · | · | · | · | 2, | · | · | · | · | · | · | · | · | · | · | · | 5, | · | · | · | · | · | · | · | · | · |
| P5 | · | · | · | · | · | · | · | · | · | 2, | · | · | · | · | · | · | · | · | · | · | · | · | · | · | · | · | · | · | · | · | · | · | · | · | · | 1, | · | · | · | · | · | · | · | · | · | · | · | · | · | · |
| P6 | · | · | · | · | · | · | · | · | 4, | · | · | · | 3, | · | · | · | · | 1, | · | · | · | · | · | · | · | · | · | · | · | · | · | · | · | · | · | · | · | · | 2, | · | · | · | · | · | · | · | · | · | · | · |
| P7 | · | · | · | · | · | · | · | · | · | 1, | · | · | · | · | · | · | · | · | · | · | · | · | · | · | 2, | · | · | · | · | · | · | · | · | · | · | · | · | · | · | · | · | · | · | · | · | · | · | · | · | · |
| P8 | · | · | · | · | · | · | · | · | · | · | · | · | · | · | · | · | · | · | · | · | · | · | · | · | · | · | · | · | · | · | · | · | · | · | · | · | 1, | · | · | · | · | · | · | · | · | · | · | · | · | · |
| P9 | · | · | · | · | · | · | · | · | · | 4, | · | · | · | · | · | 3, | · | · | · | · | · | · | · | · | · | · | · | · | · | · | · | 1, | · | · | · | · | · | · | · | · | · | · | · | · | · | · | · | · | · | 2, |
| P10 | · | 2, | 4, | · | · | · | · | · | · | · | · | · | · | · | · | · | · | · | 5, | · | · | · | · | · | · | · | · | · | · | · | · | · | · | 1, | · | · | · | · | · | · | · | · | 3, | · | · | · | · | · | · | · |
| P11 | · | · | · | · | · | 5, | · | · | · | · | · | · | · | · | · | · | · | · | · | 3, | · | · | 2, | · | · | · | · | · | · | · | · | · | 4, | · | 1, | · | · | · | · | · | · | · | · | · | · | · | · | · | · | · |
| P12 | · | · | · | · | · | · | · | · | · | · | · | · | · | · | · | · | · | 1, | · | · | · | · | · | · | · | · | · | · | · | · | · | · | · | · | · | · | · | · | · | · | · | · | · | · | · | · | · | · | 2, | · |
| P13 | · | · | · | · | · | · | · | · | 1, | · | · | · | · | · | · | · | · | · | · | · | · | · | · | · | · | · | · | · | · | · | · | · | · | · | · | · | · | · | · | · | · | · | · | · | · | · | · | · | · | · |
| P14 | · | · | · | · | · | · | · | · | · | · | · | · | · | · | · | · | · | · | · | 4, | · | · | 2, | · | · | · | · | · | · | · | 3, | · | · | · | · | 1, | · | · | · | · | · | · | · | · | · | · | · | · | · | · |
| P15 | · | · | 4, | · | · | · | · | · | · | 3, | · | · | 2, | · | · | · | · | · | · | · | · | · | · | · | · | · | · | · | · | · | · | · | · | · | · | · | 1, | · | · | · | · | · | · | · | · | · | · | · | 5, | · |
| P16 | · | · | 1, | · | · | · | · | · | · | · | · | · | · | · | · | · | · | · | · | · | · | · | · | · | · | · | · | · | · | · | 2, | · | 3, | · | · | · | · | · | · | · | · | · | · | · | · | · | · | · | · | · |
| P17 | · | · | · | · | 1, | · | · | · | · | · | · | · | · | · | · | · | · | · | · | · | · | · | · | · | · | · | · | · | · | · | · | · | · | · | · | · | · | · | · | · | · | · | · | · | · | · | · | · | · | · |
| P18 | · | · | · | · | · | · | · | · | · | · | 1, | · | · | · | · | · | · | · | · | · | 3, | · | · | 2, | · | · | · | · | · | · | · | · | · | · | · | · | · | · | · | · | · | · | · | · | · | · | · | · | · | · |
| P19 | · | · | · | · | · | 2, | · | · | · | · | · | · | · | · | · | · | · | · | · | · | · | · | · | · | · | · | · | · | · | · | · | · | 1, | · | · | · | · | · | · | · | · | · | · | · | · | · | · | · | · | · |
| P20 | · | · | · | · | · | · | · | · | · | · | · | · | · | 1, | · | · | · | 2, | · | · | · | · | · | · | · | · | · | · | · | · | · | · | · | · | · | · | · | · | · | 3, | · | · | · | · | · | · | · | · | · | · |
| P21 | · | · | · | · | · | · | · | 4, | · | · | · | · | · | · | · | · | · | · | · | · | · | · | · | · | 3, | · | 2, | · | · | · | · | · | · | · | · | · | · | · | · | · | · | · | 1, | · | · | · | · | · | · | · |
| P22 | · | · | · | · | · | · | · | 1, | · | · | · | · | · | · | · | · | · | · | · | · | · | · | · | · | · | · | · | · | · | · | · | · | · | · | · | · | · | · | · | · | · | · | · | · | · | · | · | · | 2, | · |
| P23 | · | · | · | · | · | · | · | 2, | · | · | · | · | · | · | · | · | · | · | · | · | · | · | · | · | · | · | · | · | 1, | · | · | · | · | · | · | · | · | · | · | · | · | · | · | · | · | · | · | · | · | · |
| P24 | · | · | · | · | · | · | · | · | 2, | · | · | · | 1, | · | · | · | · | · | · | · | · | · | 4, | · | · | · | · | · | · | · | · | · | · | · | · | · | · | · | · | · | · | · | · | 3, | · | · | · | · | · | · |
| P25 | · | · | · | 1, | · | 2, | · | · | 4, | · | · | · | · | · | · | · | · | · | · | · | · | · | · | · | · | · | · | · | · | · | · | · | · | · | · | · | · | · | · | · | · | · | · | · | · | 3, | · | · | · | · |
| P26 | · | · | · | · | · | · | · | · | · | · | · | · | · | · | · | · | · | · | · | · | · | 1, | · | · | · | · | · | · | · | · | · | · | · | · | · | · | · | · | · | · | · | · | · | · | · | · | · | · | · | · |
| P27 | · | · | · | · | · | · | · | · | · | · | · | · | · | · | · | · | 1, | · | · | 5, | · | · | · | · | · | · | · | · | · | · | · | · | · | · | · | · | · | · | · | · | · | · | · | 2, | 4, | · | · | · | · | 3, |
| P28 | · | · | · | · | · | · | · | · | · | · | · | · | · | · | · | · | · | · | · | · | · | · | · | · | · | · | · | · | · | · | · | · | · | · | · | · | · | · | · | · | · | · | · | · | · | · | 1, | · | · | · |
| P29 | · | · | · | · | · | · | 3, | · | · | · | · | · | · | · | · | · | · | · | · | · | · | 1, | · | · | · | · | · | · | 2, | 4, | · | · | · | · | · | · | · | · | · | · | · | · | · | · | · | · | · | · | · | · |
| P30 | · | · | · | · | · | · | · | · | · | · | · | · | 2, | · | · | · | · | · | · | · | · | · | · | · | · | · | · | · | · | · | · | · | · | · | · | · | · | · | · | · | · | 1, | · | · | · | 3, | · | · | · | · |
| P31 | · | · | · | · | · | · | · | · | · | · | 1, | · | · | · | · | · | · | · | 2, | · | · | · | · | · | · | · | · | · | · | · | · | · | · | · | · | · | · | · | · | · | · | · | · | · | · | · | · | · | · | · |
| P32 | 4, | · | · | · | · | · | · | · | · | · | · | · | · | · | · | · | · | · | · | · | · | · | · | 3, | · | · | · | · | · | 1, | · | · | · | · | · | · | · | · | · | 2, | · | · | · | · | · | · | · | · | · | · |
| P33 | · | · | · | · | · | · | · | · | · | · | · | · | · | · | · | · | · | · | · | 1, | · | · | · | · | · | · | 3, | · | · | · | · | · | · | · | · | · | · | · | 2, | · | · | · | · | · | · | · | · | · | · | · |
| P34 | · | 4, | · | · | · | · | 3, | · | · | · | · | · | · | · | · | · | · | · | · | · | · | · | · | · | · | · | · | · | · | · | · | · | · | · | · | · | 1, | · | · | · | · | · | · | · | · | 2, | · | · | · | · |
| P35 | · | · | · | · | · | · | · | · | 3, | · | · | · | · | · | · | · | · | · | · | · | · | · | · | · | · | · | · | · | · | · | 1, | · | 2, | · | 4, | · | · | · | · | · | · | · | · | · | · | · | · | · | · | · |
| P36 | · | · | · | · | · | · | · | · | · | · | · | · | · | · | · | · | · | · | 2, | · | · | · | · | · | · | · | · | · | · | · | · | 1, | · | · | · | · | · | · | · | · | · | · | · | 4, | 3, | · | · | · | · | · |
| P37 | · | 1, | · | · | · | · | · | · | · | · | · | · | · | · | · | · | · | · | · | · | · | · | · | · | · | · | · | · | · | · | · | · | · | · | · | · | · | · | · | · | · | · | · | · | · | · | · | · | · | · |
| P38 | · | · | · | · | · | · | · | · | · | · | · | · | · | · | · | · | · | · | · | · | · | · | · | · | · | · | 1, | · | · | · | · | · | · | · | · | · | · | · | · | · | · | · | · | · | · | · | · | · | · | · |
| P39 | · | · | · | · | 1, | · | · | · | · | · | · | · | · | · | · | · | · | · | · | · | · | · | 3, | · | · | · | · | · | · | · | · | · | · | · | · | · | · | · | · | 2, | · | · | · | · | · | · | · | · | · | · |
| P40 | · | · | · | · | · | · | · | · | · | · | · | · | · | · | · | · | · | · | · | · | · | · | · | · | · | · | · | · | · | · | 1, | · | · | · | · | · | · | · | · | · | · | · | · | · | · | · | · | · | · | · |
| P41 | · | · | · | · | · | · | · | · | · | · | · | · | · | · | · | · | · | · | · | · | · | · | · | · | · | · | · | · | · | · | · | · | · | · | · | · | · | · | 1, | · | · | · | · | · | · | · | · | · | · | · |
| P42 | · | · | · | · | 2, | · | · | · | · | · | · | · | · | · | · | · | · | · | 1, | · | · | · | · | · | · | · | · | · | · | · | · | · | · | · | · | · | · | · | · | · | · | · | · | · | · | · | · | · | · | · |
| P43 | · | · | · | · | · | · | · | · | · | · | · | · | · | · | · | · | · | · | · | · | · | · | · | · | · | · | · | · | · | · | · | · | · | · | · | · | · | · | · | · | 1, | · | · | · | · | · | 2, | · | · | · |
| P44 | · | · | · | · | · | · | · | · | · | · | · | · | · | · | · | · | · | · | · | · | 4, | · | · | · | · | · | · | 1, | · | · | · | · | · | 3, | · | · | · | 5, | · | · | · | · | · | · | · | · | · | · | 2, | · |
| P45 | 3, | · | 5, | · | · | 1, | · | · | · | · | · | · | · | · | · | · | · | · | 2, | · | · | · | · | · | · | · | 4, | · | · | · | · | · | · | · | · | · | · | · | · | · | · | · | · | · | · | · | · | · | · | · |
| P46 | · | · | · | · | · | · | · | · | · | · | · | · | · | · | · | · | · | · | · | · | · | · | · | · | 1, | · | · | · | · | · | · | · | · | · | · | · | · | · | · | · | · | · | · | · | · | · | · | · | · | · |
| P47 | · | · | · | · | · | 1, | · | · | · | · | · | · | · | · | · | · | · | · | · | · | · | · | · | · | · | · | · | · | · | · | · | · | · | · | · | · | · | · | · | · | · | · | · | · | · | · | · | · | · | · |
| P48 | · | · | · | · | · | · | · | · | · | · | · | · | · | 4, | · | · | · | · | · | · | · | · | · | · | · | · | · | · | · | · | · | · | · | · | 2, | · | · | · | · | · | · | · | · | · | · | 1, | · | · | · | 3, |
| P49 | · | · | · | · | · | 3, | · | · | · | · | · | · | · | · | · | · | · | · | · | · | 5, | · | 4, | · | · | · | · | · | · | 2, | · | · | · | · | · | · | · | · | · | · | · | · | 1, | · | · | · | · | · | · | · |
| P50 | · | · | · | · | · | · | 1, | · | · | · | · | · | 2, | · | · | · | · | · | · | · | · | · | · | · | · | · | · | · | · | · | · | · | · | · | · | · | · | · | · | · | · | · | · | · | · | · | · | · | · | · |
| P51 | · | · | · | · | · | · | · | · | · | · | · | · | · | · | · | · | · | 1, | · | · | · | · | · | · | · | · | · | · | · | · | · | · | · | · | · | · | · | · | · | · | · | · | · | · | · | · | · | · | · | · |
| P52 | · | · | · | · | · | · | · | · | · | 2, | · | 4, | · | · | · | · | · | · | · | · | · | · | · | · | · | · | · | · | · | · | · | · | · | · | · | · | · | · | · | 3, | · | · | · | 1, | · | · | · | · | · | · |
| P53 | · | · | · | · | · | · | · | · | · | · | · | · | · | · | · | · | · | · | · | · | · | · | · | · | · | · | · | · | · | · | · | 1, | · | · | · | · | · | · | · | · | · | · | · | · | · | · | · | · | · | · |
| P54 | · | · | · | · | · | · | 1, | · | · | · | · | · | · | · | · | · | · | · | · | · | · | · | · | · | · | · | · | · | · | · | · | · | · | · | · | · | · | · | · | · | · | · | · | · | · | · | · | · | · | · |
| P55 | · | · | · | · | · | · | · | 1, | · | · | · | · | · | · | · | · | · | · | · | · | · | · | · | · | · | · | · | · | · | · | · | · | · | · | · | · | · | · | · | · | · | · | · | · | · | · | · | · | · | · |
| P56 | · | · | · | · | · | · | · | · | · | · | · | · | · | · | · | · | 1, | · | · | · | · | · | · | · | · | · | · | · | · | · | · | · | · | · | · | · | · | · | · | · | · | · | · | 2, | · | · | · | · | · | · |
| P57 | · | · | · | · | · | · | · | · | · | 3, | · | · | · | · | · | · | · | 1, | 2, | · | · | · | · | · | · | · | · | · | · | · | · | · | · | · | · | · | · | · | · | · | · | · | 4, | · | · | · | · | · | · | · |
| P58 | · | · | · | · | 1, | · | · | · | · | · | · | · | · | 2, | · | · | · | · | · | · | · | · | · | · | · | · | · | · | · | · | · | · | · | · | · | · | · | · | · | · | · | · | · | · | · | · | · | · | · | · |
| P59 | · | · | · | · | · | · | · | · | · | · | · | · | · | · | · | · | · | 3, | · | · | 1, | · | · | · | · | · | · | · | · | · | · | 2,4, | · | · | · | · | · | · | · | · | · | · | · | · | · | · | · | · | · | · |
| P60 | · | · | · | · | · | · | · | · | · | · | · | · | · | · | · | · | 1, | · | · | · | · | · | · | · | · | · | · | · | · | · | · | · | · | · | · | · | · | · | · | · | · | · | · | · | · | · | · | · | · | · |
| P61 | · | · | · | · | · | · | · | 1, | · | · | · | · | · | · | · | · | · | · | · | · | 3, | · | · | · | · | · | · | · | · | · | · | · | · | · | · | · | · | · | · | · | · | · | · | · | 2, | 4, | · | · | · | · |
| P62 | · | · | · | · | · | · | · | · | · | · | · | · | · | · | · | · | · | · | · | · | · | · | · | · | · | · | · | · | · | · | · | · | · | · | · | · | · | 2, | · | · | · | · | · | · | · | · | 3, | · | 1, | · |
| P63 | · | · | · | · | · | · | · | · | · | · | · | · | · | 2, | · | · | · | · | · | · | · | 1, | · | · | · | · | · | · | · | · | · | · | · | · | · | · | · | · | · | · | · | · | · | · | · | · | · | · | · | · |
| P64 | · | · | · | · | · | 1, | · | · | · | · | · | · | · | · | · | · | · | · | · | · | · | · | · | · | · | · | · | · | · | · | 2, | · | · | · | · | · | · | · | · | · | · | · | · | · | · | · | · | · | · | · |
| P65 | · | · | · | · | · | · | · | · | · | · | · | · | · | · | · | · | · | · | · | · | · | · | · | · | · | · | · | · | · | · | · | · | · | · | · | · | · | · | · | · | · | · | · | 1, | · | · | · | 2, | · | · |
| P66 | · | · | · | · | · | · | · | · | · | · | · | · | · | · | · | · | · | · | · | · | · | · | · | · | · | · | · | · | · | · | · | 1, | · | · | · | · | · | · | · | · | · | · | · | · | · | · | · | · | · | · |
| P67 | · | · | 1, | · | · | · | · | · | · | · | · | · | · | · | · | · | · | · | · | · | · | · | · | · | · | · | · | · | · | · | · | · | · | · | 2, | · | · | · | · | · | · | · | · | · | · | · | · | · | · | · |
| P68 | · | · | · | · | · | · | · | · | · | · | · | · | · | · | · | · | · | · | · | · | · | · | · | · | · | · | · | · | · | · | · | · | · | · | · | · | · | · | · | · | · | · | · | · | · | · | · | 1, | · | · |
| P69 | · | · | · | · | · | · | · | · | · | · | · | · | · | · | · | · | · | · | · | 1, | · | · | · | · | · | · | · | · | · | · | · | · | · | · | · | · | · | · | · | · | · | · | · | · | · | · | · | · | · | · |
| P70 | · | · | · | · | · | · | · | · | · | · | · | 2, | · | · | · | · | · | 1, | · | · | · | · | · | · | · | · | · | · | · | · | · | · | · | · | · | · | · | · | · | · | · | · | · | · | · | 3, | · | · | · | · |
| P71 | · | · | · | · | · | · | · | · | · | · | · | · | · | · | · | · | 1, | · | · | · | · | · | · | · | · | · | · | · | · | · | · | · | · | · | · | · | · | · | · | · | · | · | · | · | · | · | · | · | · | · |
| P72 | · | · | · | · | · | · | · | · | · | · | · | · | · | · | · | · | · | · | · | · | · | · | · | · | · | · | · | · | · | · | · | · | 2, | · | · | 3, | · | · | · | · | · | · | · | · | 5, | · | · | · | 4, | 1, |
| P73 | · | · | · | · | · | · | · | · | · | · | · | · | · | · | · | · | · | · | · | · | · | · | · | · | · | · | · | · | · | 3, | · | 1, | · | · | · | · | · | 4, | · | · | · | · | · | · | · | · | 2, | · | · | · |
| P74 | · | · | · | · | · | · | · | 2, | · | · | · | · | · | · | · | · | · | · | · | · | · | · | 4, | 1, | · | · | · | · | · | · | · | · | · | · | · | · | · | · | 3, | · | · | · | · | · | · | · | · | · | · | · |
| P75 | 2, | · | · | · | · | · | · | · | · | · | · | · | · | · | · | · | 3, | · | · | · | · | · | · | · | · | · | · | · | · | · | · | · | · | · | · | · | · | · | · | · | · | · | 1, | · | · | · | · | · | · | · |
| P76 | · | · | · | · | · | · | · | · | · | · | · | · | · | · | · | · | · | · | · | · | · | · | · | · | · | · | · | · | · | · | 1, | · | · | · | · | · | · | · | · | · | · | · | · | · | · | · | · | · | · | · |
| P77 | · | · | · | · | 2, | · | · | · | · | · | · | · | · | · | · | · | · | · | · | · | · | · | · | · | · | · | · | · | · | · | 1, | · | · | · | · | · | · | · | · | · | · | · | · | · | · | · | · | · | · | · |
| P78 | · | · | · | · | · | · | 1,4 | · | · | · | · | · | · | · | · | · | · | · | · | · | · | 3, | · | · | · | · | · | · | · | · | · | · | · | · | · | · | · | 5, | · | · | · | · | · | 2, | · | · | · | · | · | · |
| P79 | · | · | · | · | · | · | · | 4, | · | · | · | · | · | · | · | · | · | · | · | · | 1, | · | · | · | · | · | · | · | · | · | · | · | · | 3, | · | · | · | · | · | · | · | · | 5, | · | · | · | · | · | 2, | · |
| P80 | 1, | · | · | · | · | · | · | · | · | · | · | · | · | · | · | · | · | · | · | · | · | · | · | · | · | · | · | · | · | · | · | · | · | · | · | · | · | · | · | · | · | · | · | · | · | · | · | · | · | · |
| P81 | · | 2, | · | · | · | · | · | · | · | · | · | · | · | · | · | · | · | · | · | · | · | · | · | · | · | · | · | · | · | · | · | · | · | 1, | · | · | · | · | · | · | · | · | · | · | · | · | 3, | · | · | · |
| P82 | · | · | · | · | · | · | 2, | · | · | · | · | · | · | · | · | · | · | · | · | · | · | · | · | · | · | · | · | · | · | · | · | · | · | · | · | · | · | 1, | · | · | · | · | · | · | · | · | · | 3, | · | · |
| P83 | 3, | · | · | · | · | · | · | · | · | · | · | · | · | · | 4, | · | · | · | · | · | · | · | · | · | · | · | · | 1, | · | · | · | · | 2, | · | · | · | · | · | 5, | · | · | · | · | · | · | · | · | · | · | · |
| P84 | · | · | · | · | · | · | · | · | · | · | · | · | · | · | · | · | · | · | · | · | · | · | · | · | · | · | · | · | · | · | · | · | · | · | · | · | · | · | · | · | · | · | · | · | 1, | · | · | · | · | · |
| P85 | · | · | 2, | · | · | · | · | · | · | · | · | 5, | · | · | · | · | · | · | · | · | · | · | · | · | · | · | · | · | · | · | · | · | · | · | 4, | · | · | 3, | · | · | · | · | · | · | · | · | · | · | · | 1, |
| P86 | · | · | · | · | · | · | · | · | · | · | · | · | 3, | · | · | · | · | · | · | · | · | · | · | · | · | · | · | · | · | · | · | · | · | · | 2, | · | · | · | · | · | · | 1, | · | · | · | 4, | · | · | · | · |
| P87 | · | · | · | 2, | · | · | · | · | · | · | · | · | · | · | · | · | · | · | · | · | · | · | · | · | · | · | · | · | · | 4, | · | · | 3, | · | · | · | · | · | · | · | · | · | 1, | · | · | · | · | · | · | · |
| P88 | · | 2, | · | · | · | · | · | 1, | · | · | · | 3, | · | · | · | · | · | · | · | · | · | · | · | · | · | · | · | · | · | · | · | · | · | · | · | · | · | · | · | · | · | · | · | · | · | · | · | · | · | · |
| P89 | · | · | · | · | · | · | · | · | · | · | · | · | · | · | · | · | · | · | · | · | · | · | · | · | · | 1, | · | 2, | · | · | · | · | · | · | · | · | · | · | · | · | · | · | · | · | · | · | · | · | · | · |
| P90 | · | · | 3, | · | · | · | · | · | · | · | · | · | · | · | · | · | · | · | · | · | · | 1, | 4, | · | · | · | · | · | 2, | · | · | · | · | · | · | · | · | · | · | · | · | · | · | · | · | · | · | · | · | · |
| P91 | · | · | · | · | · | · | · | 1, | · | · | · | · | · | · | · | · | · | · | · | · | · | · | · | · | 3, | · | · | · | · | · | · | · | · | · | · | · | · | · | · | · | · | 2, | · | · | · | · | · | · | · | · |
| P92 | · | · | · | · | · | · | · | · | 1, | · | · | · | · | · | · | · | · | · | · | · | · | · | · | · | · | · | 2, | 4, | · | 3, | · | · | · | · | · | · | · | · | · | 5, | · | · | · | · | · | · | · | · | · | · |
| P93 | · | · | · | 2, | · | · | · | · | · | · | · | · | · | · | · | · | · | · | · | · | · | · | · | 1, | · | · | · | · | · | · | · | · | · | · | · | · | · | · | · | · | · | · | · | · | · | · | · | · | · | · |
| P94 | · | · | · | · | · | · | · | · | · | · | 1, | · | · | · | · | · | · | · | · | · | · | · | · | · | · | · | · | · | · | · | · | · | · | · | · | · | · | · | · | · | · | · | · | · | · | · | · | · | · | · |
| P95 | · | 1, | · | · | · | · | · | · | · | · | · | · | · | · | · | · | · | · | · | · | · | · | · | · | · | · | · | · | · | · | · | · | · | · | · | · | · | · | · | · | · | · | · | · | 3, | · | · | · | 2, | · |
| P96 | · | 2, | · | · | · | · | · | 1, | · | · | · | · | · | · | · | · | · | · | · | · | · | · | 3, | · | · | · | · | · | · | · | · | · | · | · | · | · | · | · | · | · | · | · | · | · | · | · | · | · | · | · |
| P97 | · | · | 1, | · | · | · | 2, | · | · | · | · | · | · | · | · | · | · | · | · | · | · | · | · | · | · | · | · | · | · | · | · | · | 5, | 3, | · | · | · | · | · | · | · | · | · | · | · | 4, | · | · | · | · |
| P98 | · | · | · | · | · | · | · | · | · | · | · | 3, | · | · | 5, | · | · | · | · | 4, | · | · | · | 2, | · | · | · | · | · | · | · | · | · | · | · | · | · | · | · | · | · | · | 1, | · | · | · | · | · | · | · |
| P99 | · | · | · | · | · | · | · | · | · | · | · | · | · | · | · | · | · | · | · | · | · | · | · | · | · | · | · | · | · | · | · | · | · | · | · | · | · | 1, | · | 2, | · | · | · | · | · | · | · | 3, | · | · |
| P100 | · | · | · | · | · | · | · | · | · | · | · | · | 2, | · | · | · | · | · | · | · | · | · | · | · | · | · | · | · | · | · | · | · | · | · | · | · | · | · | · | · | · | · | · | · | 1, | · | · | · | · | · |